\documentclass[twocolumn,prb,preprintnumbers,amsmath,amssymb,unsortedaddress]{revtex4-1}

\usepackage{graphicx}
\usepackage{dcolumn}
\usepackage{bm}

\begin{document}


\title{Atomistic Boron-Doped Graphene Field Effect Transistors:\\
A Route towards Unipolar Characteristics}

\author{Paolo Marconcini}
\affiliation
{Dipartimento di Ingegneria dell'Informazione, 
Universit\`a di Pisa,
Via Caruso 16, 56122 Pisa, Italy}
\author{Alessandro Cresti}
\affiliation
{IMEP-LAHC, UMR 5130 (Grenoble INP/UJF/CNRS/Universit\'e de Savoie),
Minatec, 3 Parvis Louis N\'eel, 
38016 Grenoble, France}
\author{Fran\c{c}ois Triozon}
\affiliation
{CEA LETI-MINATEC, 17 rue des Martyrs, 38054 Grenoble, France}
\author{Gianluca Fiori}
\affiliation
{Dipartimento di Ingegneria dell'Informazione, 
Universit\`a di Pisa,
Via Caruso 16, 56122 Pisa, Italy}
\author{Blanca Biel}
\affiliation
{Dpto. Electr\'onica y Tecnolog{\'\i}a de Computadores, Universidad
de Granada, Facultad de
Ciencias, Campus de Fuente Nueva, and CITIC, Campus de Aynadamar,
Universidad de Granada E-18071 Granada, Spain}
\author{Yann-Michel Niquet}
\affiliation
{L\_Sim, SP2M, UMR-E CEA/UJF-Grenoble 1, INAC, Grenoble,
France}
\author{Massimo Macucci}
\affiliation
{Dipartimento di Ingegneria dell'Informazione,
Universit\`a di Pisa,
Via Caruso 16, 56122 Pisa, Italy}
\email{m.macucci@mercurio.iet.unipi.it}
\author{Stephan Roche}
\affiliation
{CIN2 (ICN-CSIC) and Universitat Aut\'{o}noma de Barcelona, Catalan
Institute of Nanotechnology, Campus UAB, 08193 Bellaterra (Barcelona),
Spain, and
ICREA, Instituci\'{o} Catalana de Recerca i Estudis Avan\c{c}ats,
08070 Barcelona, Spain}

\begin{abstract}
We report fully quantum simulations of realistic models of boron-doped
graphene-based field effect transistors, including atomistic details
based on DFT calculations. We show that the self-consistent solution
of the three-dimensional (3D) Poisson and Schr\"odinger equations
with a representation in terms of a tight-binding Hamiltonian manages to
accurately reproduce the DFT results for an isolated boron-doped graphene
nanoribbon. Using a 3D  Poisson/Schr\"odinger solver within the
Non-Equilibrium Green's Functions (NEGF) formalism, self-consistent
calculations of the gate-screened scattering potentials induced by the
boron impurities have been performed, allowing the theoretical
exploration of the tunability of transistor characteristics. The
boron-doped graphene transistors are found to approach unipolar behavior
as the boron concentration is increased, and by tuning
the density of chemical dopants the electron-hole transport asymmetry
can be finely adjusted. Correspondingly, the onset of a
mobility gap in the device is observed. Although the 
computed asymmetries are not sufficient to warrant proper device operation, 
our results represent an initial step in the direction of improved 
transfer characteristics and, in particular, the developed simulation 
strategy is a powerful new tool for modeling doped graphene nanostructures.
\end{abstract}

\maketitle

The discovery of graphene has opened a promising alternative to
silicon-based electronics.~\cite{Novoselov1, Novoselov2, GFET}
Reported charge mobilities in undoped graphene layers are actually
orders of magnitude larger than those measured in silicon, 
but the unfortunate zero-gap semiconductor nature of this material
severely limits the achievable $I_{\rm on}$/$I_{\rm off}$ 
ratio~\cite{ITRS} (ratio ---infinite for an ideal switch---
of the current flowing through the device in the conducting state to
that flowing in the nonconducting state) for graphene transistors, making
them not yet able to compete with mainstream silicon
technologies.~\cite{Iannaccone}
Another adverse feature of graphene-based devices is their ambipolar
electrical behavior, which precludes the development of a complementary
logic architecture.

A way to induce an energy gap in graphene is to reduce
its lateral dimension.~\cite{Nakada,Brey,CastroNeto,Marconcini}
By using e-beam lithographic techniques and oxygen plasma etching processes,
graphene nanoribbons (GNRs) down to a width of 10 nanometers can be
successfully fabricated.~\cite{GNREXPT1,GNREXPT2} The resulting energy gap, 
however, is not sufficient for the achievement of a practically useful 
$I_{\rm on}$/$I_{\rm off}$ ratio in a transistor structure. Chemical, 
in-solution techniques have been developed that allow one to obtain nanoribbons
with a smaller width~\cite{GNREXPT3,GNREXPT4} but are hard to integrate
into a large-scale device production technology. A mixed approach has been
followed by Wang and Dai,~\cite{wangdai} who use lithography to define the
nanoribbons, whose width is then shrunk by means of gas-phase chemical
etching. Although quite interesting, this procedure could have some limitation
in terms of the achievable device density. Furthermore,
energy bandgaps are very unstable with regards to edge
reconstruction and defects,~\cite{Cresti} thus preventing the fabrication 
of graphene transistors with performances comparable with their silicon 
counterparts. Indeed, almost
ideal graphene nanoribbons have been obtained by means of careful chemical 
synthesis,~\cite{natureatomic} however such an approach cannot be applied to 
the fabrication of complex circuits on insulating substrates in a
straightforward way.
Strong chemical damage of graphene could in principle be used to tune the 
current
flow, but with the drawback of a significant decay of the 
conductance that leads to poor performance in terms of current 
drive.~\cite{Mobgapsepoxide1,
Mobgapsepoxide2,Mobgapsepoxide3,Mobgapsepoxide4,Bruzzone}

Some of us have recently proposed a new class of devices based on the
chemical doping of graphene by boron or nitrogen.~\cite{Biela,Bielb1,Bielb2}
Using first-principles and mesoscopic quantum transport calculations
(within the Landauer-B\"uttiker formalism) some indications of different
behaviors for electrons and holes were predicted in ribbons with lateral
sizes as large as 10~nm, with mobility gaps in the order of
1~eV.~\cite{Biela,Bielb1} Here the mobility gaps are
typically defined by the corresponding energy window for which the
conductance is smaller than a typical value of $G=10^{-2}G_{0}$
($G_{0}=2e^{2}/h)$ (see Ref.~\cite{Crestinew} for details).
Recent experimental findings confirmed the possibility
to include boron and nitrogen impurities in graphene,~\cite{KimN1,KimN2,KimN3}
therefore suggesting doped graphene transistors as a genuine alternative
to ultranarrow GNRs with a more controlled tuning of the conductance
features. However, fully self-consistent simulations
of doped graphene transistors are mandatory to ascertain the real impact
of graphene doping.

We explore the effect, at room temperature, of boron doping in graphene-based
field effect transistors using the device geometry illustrated in
Fig.~\ref{fig1}. Self-consistent NEGF calculations are performed, allowing
for a proper treatment of the screened impurity potential and a more
robust assessment of the doping effects on the electron-hole transport
asymmetry and on the onset of mobility gaps.

\section{Results and discussion}

Parameters for our calculations have been obtained from a first-principle
approach, as we detail in the following.

As an initial step, first-principle density functional theory 
(DFT) calculations
were undertaken for a single substitutional boron impurity in two-dimensional
(2D) graphene and in GNRs,~\cite{Biela,Bielb1}
by means of the SIESTA code,~\cite{SIESTA} within
the local density approximation and using a double-$\zeta$ basis set.
Following the methodology in Refs.~\cite{Bielb1,Adessi}, the onsite energies
of the $p_z$ orbitals for the (bulk) 2D case were extracted. Their
variation with the distance from the boron atom is shown in Fig.~\ref{fig2}
(inset). The boron atom acts as an electron acceptor: the resulting
impurity potential is repulsive for electrons and the screening length
is found to be of the order of a few angstroms
(the screening length~\cite{Stern} is defined as the
distance from the charged impurity at which the surrounding mobile
charges, rearranging under the electrostatic action of
the impurity, reduce its potential by a factor equal to Euler's number $e$).

The transport properties
of ``quasi-metallic'' and semiconducting armchair GNRs 
(following Ref.~\cite{Son},
we refer to an armchair GNR with $N$ dimers contained in its unit cell
as an ``N-aGNR'') with a single boron impurity
at different sites across the ribbon width are then computed, using
the full {\it ab initio} Hamiltonian, in the Landauer-B\"uttiker
framework.~\cite{Biela} Boron impurities actually induce a strong hole
backscattering, with much weaker effect on the electron side. A similar
result was already obtained in boron-doped carbon
nanotubes,~\cite{Cohen1,Cohen2}
and attributed to strong backscattering at the resonance energies of
the quasibound states localized around the boron impurity.
Due to the lack of the rotational symmetry present in
carbon nanotubes, in GNRs the
energy of the conductance dip on the hole side depends on the position
of the impurity with respect to the ribbon edge (as illustrated in
Fig.~\ref{fig2} and Fig.~\ref{fig3}).  

In Ref.~\cite{Bielb1} it has been shown that these {\it ab initio}
results can be fairly well reproduced with a simpler tight-binding model
with a single orbital per carbon and boron atom.
For each atom, the onsite energy is taken as the impurity 
potential extracted from the {\it ab initio} calculation. The onsite energy
of the boron atom is the only parameter that has to be adjusted in order to
achieve the best match between tight-binding and {\it ab initio} transmission
profiles. The success of such a simple
tight-binding model in describing boron impurities originates from the
preservation of the $sp^2$ hybridization of graphene in the
presence of substitutional boron.

By using this tight-binding model, transport studies in large armchair
GNRs with a random distribution of boron atoms, with concentrations
between $0.02$\%  and $0.2$\%, have been performed in Ref.~\cite{Bielb1}{.}
The addition of the transmission dips induced by all impurities leads to a
{\it mobility gap} in the valence band, {\it i.e.} an energy interval where
conductance is nearly suppressed by scattering, while transport in the
conduction band is preserved (the opposite behavior would be found in
the case of nitrogen doping). This occurs for
semiconducting as well as for ``quasi-metallic'' GNRs, independently of ribbon
width, as long as the potential can be assumed to be constant along the
channel. The mobility gap increases with boron concentration and ribbon
length. Ref.~\cite{Bielb1} suggested a possible exploitation of 
this phenomenon to obtain field-effect transistors (FET) with wide GNRs ($>
10$ nm), within the reach of conventional lithography. The absence of
a significant bandgap in these ribbons could be compensated for by the
mobility gap. However, the above mentioned study has been performed for
ribbons without any gate-induced charge in the
channel, since the impurity potential considered was the one extracted
directly from an {\it ab initio} calculation without taking into account
the effects of screening when a gate is applied.  

\begin{figure}[!ht]
\begin{center}
\includegraphics[width=0.9\columnwidth]{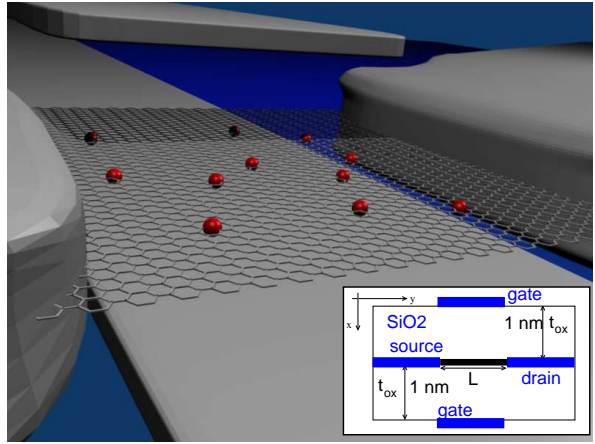}
\end{center}
\caption{Sketch of the simulated double gate GNR field effect transistor
(FET): red dots represent the substitutional boron atoms. In the inset, the
longitudinal cross-section of the transistor is shown 
(the scale is different along the $x$ and $y$ directions, to make the drawing 
clearer).}
\label{fig1}
\end{figure}

In order to simulate the behavior of a realistic transistor structure,
we need to perform a self-consistent calculation to include
the effects of an applied gate voltage and of a non-zero drain-source bias
on the charge density. This can be done, at a reasonable
level of computational complexity, with a tight-binding model set up 
with parameters properly extracted from the {\it ab initio} results.
However, differently from Ref.~\cite{Bielb1}, the impurity
potential obtained from {\it ab initio} approaches cannot be directly used in
a self-consistent simulation, since it would be overscreened, as a result of 
the Poisson/Schr\"odinger iterations. The approach chosen here is to mimic
DFT self-consistency with the tight-binding model and a proper distribution
of fixed charges; then the final impurity potential arises from the
tight-binding Poisson/Schr\"odinger self-consistency.

In particular, as we will detail in the Methods section, the
{\it ab initio} results can be reproduced considering a fixed charge $+e$
(where $e$ is the elementary charge) in correspondence of each carbon atom,
and using the following tight-binding representation: null onsite energies are 
considered in correspondence of all atoms, while, following 
Ref.~\cite{Son}, 
the hopping parameter is equal
to $t_p=-2.7$ eV for all the pairs of nearest neighbor atoms, with the
exception of the edge dimers, for which a value of $1.12\,t_p$ is used instead.

To validate this approach, a self-consistent Poisson/Schr\"odinger calculation
(described in the Methods section) is performed for an isolated graphene layer
with a single impurity, as in the {\it ab initio} calculation. 
As shown in Fig.~\ref{fig2} (inset), the obtained self-consistent
potential around the impurity is very close to the {\it ab initio} result. 
Also the {\it ab initio} transmission spectra are accurately reproduced for
all positions of the impurity, keeping the boron onsite energy equal to zero
and with no further parameter adjustment.
Transmission spectra comparisons are shown in Fig.~\ref{fig2}.
\vskip5pt
\par\noindent
\begin{figure}[htb]  
\begin{center}
\includegraphics[width=0.9\columnwidth]{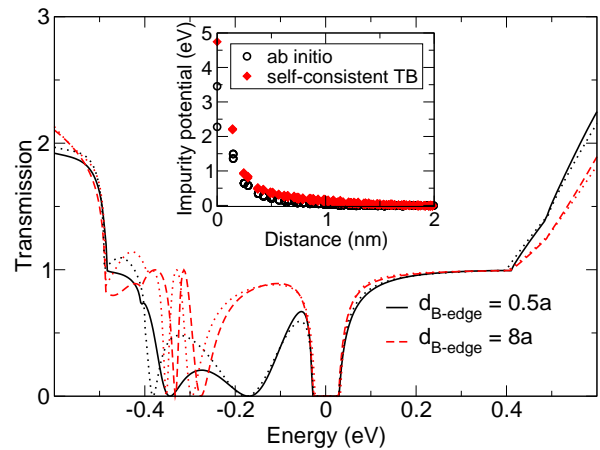}
\caption{Transmission through an isolated 32-aGNR (3.81~nm wide)
with one boron impurity
placed at two different distances from the ribbon edge: $0.5a$ and $8a$,
where $a = 2.46$~{\AA} is the hexagonal lattice constant. The bold lines
represent the results with our self-consistent tight-binding model.
The thin dotted line represents the {\it ab initio} results.
Inset: comparison between the results for the impurity potential in 2D
graphene obtained with self-consistent tight-binding and {\it ab initio} 
approaches.} 
\label{fig2}
\end{center}
\end{figure}

Before performing full NEGF calculations, the effect of a gate with a
nonzero applied voltage was analyzed on
short ribbons with a single impurity. The ribbon is surrounded by a double
gate with a dielectric thickness of 1 nm, as shown in Fig.~\ref{fig1}. 
The source-drain voltage and the Fermi level are set to $0$. The charge
and potential on the ribbon adjust self-consistently as a function of
the gate potential $V_G$. First, the calculation was performed at $V_G = 0$
with vacuum chosen as gate dielectric. The obtained potential is very
close to the result discussed above for the neutral isolated
ribbon. This gate configuration does not induce
any significant additional screening. A more realistic model is obtained
by considering a silicon dioxide dielectric between the gate plates and the 
ribbon, with relative dielectric permittivity $\varepsilon_r = 3.9$. However, 
thin vacuum layers are kept just below and above the ribbon, with a thickness
of $0.3$ nm, of the order of the calculated graphene/SiO$_2$
distance.~\cite{hossain_apl2009}
The impurity potentials and the corresponding
transmission spectra are shown in Fig.~\ref{fig3}, for $V_G = 0$ and
$V_G = -0.5$ V. Screening is enhanced at a negative gate voltage, due to
the accumulation of holes in the ribbon. This has a strong effect on
transmission: the conductance dip is less pronounced and is shifted in
energy, which could compromise the presence of a clear mobility gap in
a transistor configuration. It was verified that the effect of 
the gates is much larger than that of the thin air layers, in terms of the global
electrostatics, and, in particular, of the screening. This is the reason why
in the runs for the evaluation of the transistor characteristics we will not
introduce the air layers, that would make convergence of the
Poisson-Schr\"odinger iterations much slower.
\vskip5pt
\par\noindent
\begin{figure}[!ht]    
\begin{center}
\includegraphics[width=0.9\columnwidth]{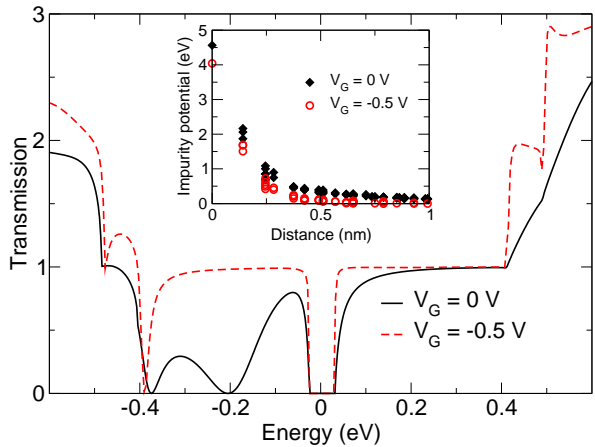}
\caption{Influence of the gate voltage on the transmission spectrum of
a 32-aGNR with one boron impurity placed at a distance $0.5a$ from the
edge. The calculation was performed with silicon dioxide as gate dielectric
and a thin vacuum layer surrounding the graphene sheet.
Inset: influence of the gate voltage on the impurity potential.}
\label{fig3}
\end{center}
\end{figure}

To assess the performance of complete boron-doped graphene transistors
biased in a typical operating point, we have
used the open-source code NanoTCAD ViDES.~\cite{NanoTCAD} In this code,
the 3D Poisson equation is solved self-consistently with the Schr\"odinger
equation (with open-boundary conditions), within the Non-Equilibrium
Green's Function (NEGF) formalism.~\cite{Datta} Schottky contacts have
been modeled at the GNR ends with the phenomenological approach described
in Ref.~\cite{Guo}, including self-energies that mimic metallic contacts.

\begin{figure}[!ht]
\begin{center}
\includegraphics[width=0.9\columnwidth]{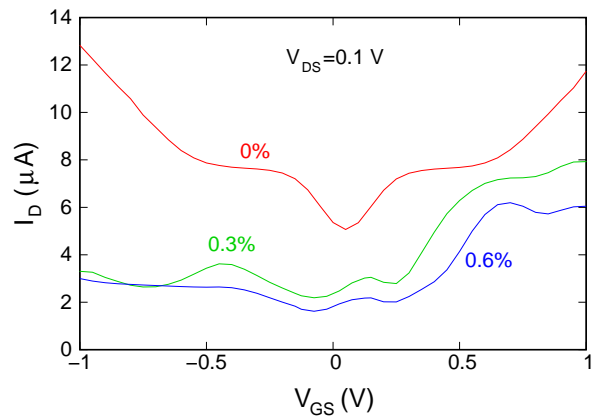}
\end{center}
\caption{Transfer characteristics for two different boron doping
concentrations (0.3\% and $0.6\%$), for a given random distribution of
the dopant atoms. The transfer characteristic for the undoped device
is also reported.} 
\label{fig4}
\end{figure}

The considered device structure is reported in the inset of 
Fig.~\ref{fig1}. The device is a double-gate FET and the channel is a
32-aGNR (3.81~nm wide)
embedded in SiO$_2$ dielectric. This is consistent with typical 
nanofabrication techniques, in which the dielectric is deposited after the 
creation of the drain and source contacts. Top and bottom field oxides are
1~nm thick, and the channel length is equal to 20~nm. In Fig.~\ref{fig4},
we show the transfer characteristics of the GNR-FET for a drain-source
voltage $V_{DS}$ = 0.1~V, for
different boron doping concentrations (defined as the ratio of the
number of boron atoms to the total number of atoms in the channel). Dopant
atoms are randomly placed along the channel. Results in Fig.~\ref{fig4}
refer to a single distribution of dopants.

As expected, transfer characteristics for the undoped device are symmetric
with respect to $V_{GS}=V_{DS}/2 = 0.05$~V. As soon as some boron doping
is considered, the p-branch of the $I_{DS}$-$V_{GS}$ curve is suppressed,
as previously observed also in carbon nanotubes.~\cite{Cohen1,Cohen2}
The p-branch suppression increases as the doping concentration
is increased.
From these results, it appears that with the considered
boron doping concentrations, while still not obtaining a sufficient
$I_{on}/I_{off}$ ratio for logic applications~\cite{ITRS} with the
considered 32-aGNR (which, however, having a ``quasi-metallic'' behavior
in the absence of doping, represents, in this regard, a worst-case
scenario), a clear onset of unipolarity occurs. 
Increasing boron doping, while further reducing the
current in the p-branch (due to holes), also leads to a degradation of
the device performance above threshold.

\begin{figure}[!ht]
\begin{center}
\includegraphics[width=0.9\columnwidth]{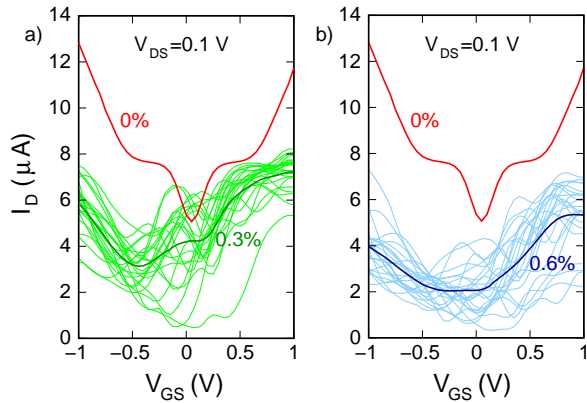}
\end{center}
\caption{Transfer characteristics for a) 0.3\% and b) 0.6\% boron doping
concentration, considering a statistics of 23 randomly distributed dopant
atoms. The thick solid
lines correspond to the average transfer characteristic.
For comparison, we report also the transfer characteristic for the
undoped device.}
\label{fig5}
\end{figure}

In Fig.~\ref{fig5}(a)-(b), we show the transfer
characteristics for 23 different distributions of doping atoms along the
channel, for the same doping concentrations as in Fig.~\ref{fig4}. The
thick solid lines correspond to the average transfer characteristics
computed over the considered statistics of the doping distribution.
As can be seen, different doping distributions lead to a large dispersion
of the transfer characteristics, which could be reduced if several nanoribbon
devices were operated in parallel (as can be necessary in some cases in order
to increase the on current).

Nevertheless, we remark that the electron-hole asymmetry is clearly
developing with chemical doping, and, based on the results of 
Ref.~\cite{Bielb1}, it can be strengthened by 
increasing the gate length up to few hundreds of
nanometers (especially for the case with $0.3\%$). Indeed, longer
channel lengths could result in cumulated contributions of multiple
scattering effects and localization, which will even more inhomogeneously
impact on transport characteristics (as shown in equilibrium
situations~\cite{Bielb1}). Due to computational limitations,
we are not reaching such channel lengths here. Computational 
hurdles, in terms of difficulties in achieving self-consistence, also limit 
the doping level that can currently be simulated with our method to 
about 0.6\%. Thus, at present we cannot handle the extremely 
high doping concentrations, of the order of 10\%, that have been reached in a 
recent experiment~\cite{tang} reporting unexpectedly large gap opening effects.
\par\noindent

\section{Conclusion}

In conclusion, by coupling a DFT-based tight-binding parameterization 
(validated {\it via} a multi-scale approach based on {\it ab initio} techniques)
with self-consistent NEGF quantum-mechanical transport simulations, we have
found the onset of an electron/hole transport asymmetry and mobility gaps in
boron-doped graphene nanoribbon transistors,
confirming that chemical doping could serve as a way to shape the
current-voltage characteristics of such devices. 
This is just a first step, because the achieved $I_{on}/I_{off}$
ratio, although for a worst-case condition, is still far too small for 
practical applications. There is however ample room for improvement and 
several different approaches could be used to selectively incorporate
chemical entities into graphene, thus engineering separate chemically
modified areas from a single graphene layer.
A doping strategy could therefore be elaborated to implement pn junctions,
logic gates and complex architectures directly patterned on the same
underlying graphene layer. Conventional state-of-the-art patterning
methods, such as e-beam lithography, could be exploited
to design superimposed material architectures (with either metallic
or other semiconducting materials) onto the selectively doped and
functionalized graphene-based substrates. 
Although our transistor simulations 
have been performed for graphene nanoribbons with a width below 5~nm (due to 
the otherwise prohibitive computational cost), we wish to point out
that the scaling analysis detailed in Refs.~\cite{Biela,Bielb1} suggests
that our main finding (electron-hole transport asymmetry) should be confirmed
for ribbon widths above 10~nm, that is within the reach of conventional
fabrication lithography techniques. Finally, longer ribbon lengths and
low temperature are also factors which are capable of widening the mobility
gaps.~\cite{Bielb1}

\section{Methods}

The simulation approach is based on the observation
that the DFT results can be recovered performing a self-consistent
Poisson/Schr\"odinger calculation, within the NEGF formalism,
on the system, which is described using a
tight-binding model and a proper distribution of fixed charges.

In detail, we found that, following this procedure, the tuning
of a single parameter of the tight-binding model is
sufficient to properly reproduce the {\it ab initio} impurity potential
and transmission spectra. The method is inspired by what actually occurs in
the DFT calculation. The total charge of a carbon nucleus and of its
core electrons is $+4e$, while it is
$+3e$ for boron. In the isolated atoms, this charge is compensated
by $4$ valence electrons for carbon and $3$ valence electrons
for boron. In graphene, $3$ electrons per atom
(including, in the case of doped graphene, the boron atoms) tend to
hybridize in a ``$sp^2$'' fashion, while the remaining valence electron
on each carbon lies on the ``$\pi$'' bonding subband formed by the $p_z$
orbitals. 
Also the $p_z$ orbital on the boron atom is not empty, since it tends
to attract an electron from the neighboring carbon atoms, and this
determines the shape of the impurity potential shown in Fig.~\ref{fig2}. 
In the single $p_z$ orbital tight-binding approach, the ``$sp^2$'' valence
electrons are included in the global ionic charge of the atoms, which
becomes $+e$ for carbon and $0$ for boron. 
Therefore this is the net charge ``localized'' on each atom 
that we have to sum to the negative charge of the $p_z$ atomic
orbitals considered in the simulation. As far as the tight-binding
parameters are concerned, the onsite energy for carbon atoms is
$0$, while the hopping energy between nearest neighbors is set to $t_p
= -2.7$ eV. The hopping energy of edge dimers is adjusted to $1.12t_p$
in order to reproduce the small bandgap obtained in DFT for ``quasi-metallic''
N-aGNRs ($N = 3p+2$, where $p$ is an integer).~\cite{Son,Fujita}

Only the onsite energy of boron may be further adjusted,
performing a self-consistent Poisson/Schr\"odinger calculation for an
isolated graphene ribbon with a single boron atom and comparing the
obtained results with those deriving from the {\it ab initio} simulation
of the same structure.

For the Poisson equation the same boundary conditions are used in the
{\it ab initio} and in the tight-binding calculations: Neumann boundary
conditions ({\it i.e.} derivative of the potential energy, and thus electric field,
constant and, in particular, zero) in the longitudinal direction, at the ends 
of the nanoribbon, and periodic in the transverse direction, with a
separation of 2~nm between the adjacent edges of nanoribbon replicas
(indeed, with a large enough separation between the replicas this is equivalent
to a Neumann boundary condition). The gate stack has been
treated considering a constant permittivity, the one of silicon oxide, between
the nanoribbon and each gate, where a Dirichlet boundary condition is assumed.
At each iteration, the eigenvalues of the
Hamiltonian are computed and the Fermi level is adjusted to obtain charge
neutrality. The local charge density is then calculated and the Poisson
equation is solved. The obtained electrostatic potential is subtracted
from the onsite energies and the process is repeated until self-consistency
is achieved.
For all of the meshes tested
in this work, it was possible
to fit the {\it ab initio} data simply by adjusting the boron 
onsite energy. A typical choice was to spread the charge of each atom in a 
spatial region of half-width $a_{CC}$, the interatomic distance, and to choose 
a mesh interval of 0.1~nm, small enough to achieve convergence of the Poisson 
solution. This charge repartition around each atom was defined using a 
smooth envelope function. In this case, the best fitting was obtained for a 
value of the boron onsite energy of about zero, thus substantially coincident 
with that for carbon.

\begin{acknowledgements}

B.B. acknowledges financial support from the Juan de la Cierva Program
and the
FIS2008-05805 Contract of the Spanish MICINN. A.C. acknowledges the
Fondation Nanosciences {\it via} 
the RTRA Dispograph project. We acknowledge the use of the software
VMD \cite{VMD2} for graphical design. 
This work was partly funded
by the European Union under Contract No. 215752 GRAND (GRAphene-based
Nanoelectronic Devices), by the French National Research Agency (ANR),
in the framework of its 2009 program in Nanosciences, Nanotechnologies 
\& Nanosystems (P3N2009), 
through the NANOSIM-GRAPHENE project n$^{\rm o}$ ANR-09-NANO-016-01,
and 
by the Italian Ministry for the University and Research (MIUR) through
the GRANFET project (n. 2008S2CLJ9). Part of the calculations were performed
with the support of the GENCI (Grand Equipment National de Calcul Intensif)
initiative.

\end{acknowledgements}


\begin{thebibliography}{50}




\bibitem{Novoselov1}
Geim,~A.~K.; Novoselov,~K.~S. 
The Rise of Graphene.
\emph{Nat. Mater.} \textbf{2007},
\emph{6}, 183--191.

\bibitem{Novoselov2}
Novoselov,~K.~S. 
Nobel Lecture: Graphene: Materials in the Flatland.
\emph{Rev. Mod. Phys.} \textbf{2011},
\emph{83}, 837--849.

\bibitem{GFET}
Schwierz,~F. 
Graphene Transistors.
\emph{Nat. Nanotechnol.} \textbf{2010}, \emph{5}, 487--496.

\bibitem{ITRS} 
ITRS roadmap 2009: http://www.itrs.net
(accessed May 28, 2012).

\bibitem{Iannaccone}
Iannaccone,~G.; Fiori~G.; Macucci~M.; Michetti~P.; Cheli~M.; Betti~A.;
Marconcini~P.
Perspectives of Graphene Nanoelectronics: Probing Technological Options
with Modeling.
In \emph{Proceedings of IEDM 2009}, Baltimora, USA,
December 7-9, 2009; IEEE Conference Proceedings: 2009; pp. 245--248.

\bibitem{Nakada} Nakada,~K.; Fujita,~M.; Dresselhaus,~G.; Dresselhaus,~M.~S.
Edge State in Graphene Ribbons: Nanometer Size Effect and Edge Shape
Dependence.
\emph{Phys. Rev. B} \textbf{1996}, \emph{54}, 17954--17961.

\bibitem{Brey} Brey,~L.; Fertig,~H.~A.
Electronic States of Graphene Nanoribbons Studied with the Dirac Equation.
\emph{Phys. Rev. B} \textbf{2006}, \emph{73}, 235411.

\bibitem{CastroNeto} Castro~Neto,~A.~H.; Guinea,~F.; Peres,~N.~M.~R.;
Novoselov,~K.~S.; Geim,~A.~K.
The Electronic Properties of Graphene.
\emph{Rev. Mod. Phys.} \textbf{2009}, \emph{81}, 109--162.

\bibitem{Marconcini} Marconcini,~P.; Macucci,~M.
The $k \cdot p$ Method and Its Application to Graphene, Carbon Nanotubes
and Graphene Nanoribbons: the Dirac Equation.
\emph{Riv. Nuovo Cimento Soc. Ital. Fis.} \textbf{2011},
\emph{34}, 489--584. See also
http://brahms.iet.unipi.it/supplem/reviewgraph.html
(accessed May 28, 2012).

\bibitem{GNREXPT1}
Han,~M.~Y.; \"Ozyilmaz,~B.; Zhang,~Y.; Kim,~Ph. 
Energy Band-Gap Engineering of Graphene Nanoribbons.
\emph{Phys. Rev. Lett.} \textbf{2007}, \emph{98}, 206805.

\bibitem{GNREXPT2}
Han,~M.~Y.; Brant,~J.~C.; Kim,~Ph. 
Electron Transport in Disordered Graphene Nanoribbons.
\emph{Phys. Rev. Lett.} \textbf{2010}, \emph{104}, 056801.

\bibitem{GNREXPT3}
Wang,~X.; Ouyang,~Y.; Li,~X.; Wang,~H.; Guo,~J.; Dai,~H.
Room-Temperature All-Semiconducting Sub-10-nm Graphene Nanoribbon
Field-Effect Transistors.
\emph{Phys. Rev. Lett.} \textbf{2008} \emph{100}, 206803.

\bibitem{GNREXPT4}
Li,~X.; Wang,~X.; Zhang,~L.; Lee,~S.; Dai,~H.
Chemically Derived, Ultrasmooth Graphene Nanoribbon Semiconductors.
\emph{Science} \textbf{2008}, \emph{319}, 1229--1232.

\bibitem{wangdai}
Wang,~X.; Dai,~H.
Etching and Narrowing of Graphene from the Edges.
\emph{Nature Chem.} \textbf{2010}, \emph{2}, 661--665.

\bibitem{Cresti}
Cresti,~A.; Nemec,~N.; Biel,~B.; Niebler,~G.; Triozon,~F.; Cuniberti,~G.;
Roche,~S. 
Charge Transport in Disordered Graphene-Based Low Dimensional Materials.
\emph{Nano Res.} \textbf{2008}, \emph{1}, 361--394.

\bibitem{natureatomic} Cai,~J.; Ruffieux,~P.; Jaafar,~R., Bieri,~M.;
Braun,~T.; Blankenburg,~S.; Muoth,~M.; Seitsonen,~A.~P.; Saleh,~M.;
Feng,~X.; {\it et al.}
Atomically Precise Bottom-Up Fabrication of Graphene Nanoribbons.
\emph{Nature} \textbf{2010}, \emph{466},
470--473.

\bibitem{Mobgapsepoxide1} 
Moser,~J.; Tao,~H.; Roche,~S.; Alzina,~F.; Sotomayor Torres,~C.~M.;
Bachtold,~A. 
Magnetotransport in Disordered Graphene Exposed to Ozone:
From Weak to Strong Localization.
\emph{Phys. Rev. B} \textbf{2010}, \emph{81}, 205445. 

\bibitem{Mobgapsepoxide2}
Leconte,~N.; Moser,~J.;
Ordej\'on,~P.; Tao,~H.; Lherbier,~A.; Bachtold,~A.; Alsina,~F.;
Sotomayor Torres,~C.~M.; Charlier,~J.~C.; Roche,~S.
Damaging Graphene with Ozone Treatment: A Chemically Tunable
Metal-Insulator Transition.
\emph{ACS Nano} \textbf{2010}, \emph{4}, 4033--4038.

\bibitem{Mobgapsepoxide3}
Leconte,~N.; Lherbier,~A.; Varchon,~F.; Ordejon,~P.; Roche,~S.;
Charlier,~J.-C.
Quantum Transport in Chemically Modified Two-Dimensional Graphene:
From Minimal Conductivity to Anderson Localization.
\emph{Phys. Rev. B} \textbf{2011}, \emph{84}, 235420. 

\bibitem{Mobgapsepoxide4}
Cresti,~A.; Lopez-Bezanilla,~A.; Ordej\'on,~P.; Roche,~S.
Oxygen Surface Functionalization of Graphene Nanoribbons for Transport
Gap Engineering.
\emph{ACS Nano} \textbf{2011}, \emph{5}, 9271--9277.

\bibitem{Bruzzone} Bruzzone~S.; Fiori~G.
{\it Ab-Initio} Simulations of Deformation Potentials and Electron Mobility in
Chemically Modified Graphene and Two-Dimensional Hexagonal Boron-Nitride.
\emph{Appl. Phys. Lett.} \textbf{2011}, \emph{99}, 222108.

\bibitem{Biela} 
Biel,~B.; Blase,~X.; Triozon,~F.; Roche,~S.
Anomalous Doping Effects on Charge Transport in Graphene Nanoribbons.
\emph{Phys. Rev. Lett.} \textbf{2009}, \emph{102}, 096803.

\bibitem{Bielb1} 
Biel,~B.; Triozon,~F.; Blase,~X.; Roche,~S.
Chemically Induced Mobility Gaps in Graphene Nanoribbons: A Route for
Upscaling Device Performances.
\emph{Nano Lett.} \textbf{2009}, \emph{9}, 2725--2729.

\bibitem{Bielb2} 
Lherbier,~A.; Biel,~B.; Niquet,~Y.-M.; Roche,~S.
Transport Length Scales in Disordered Graphene-Based Materials:
Strong Localization Regimes and Dimensionality Effects.
\emph{Phys. Rev. Lett.} \textbf{2008}, \emph{100}, 036803.

\bibitem{Crestinew}
Cresti,~A.; Roche,~S.
Range and Correlation Effects in Edge Disordered Graphene Nanoribbons.
\emph{New J. Phys.} \textbf{2009}, \emph{11}, 095004.

\bibitem{KimN1}
Zhao,~L.; He,~R.; Rim,~K.~T.; Schiros,~T.; Kim,~K.~S.; Zhou,~H.;
Guti\'errez,~C.; Chockalingam,~S.~P.; Arguello,~C.~J.; P\'alov\'a,~L.;
{\it et al.}
Visualizing Individual Nitrogen Dopants in Monolayer Graphene.
\emph{Science} \textbf{2011}, \emph{333}, 999--1003. 

\bibitem{KimN2}
Usachov,~D.; Vilkov,~O.; Gr\"uneis,~A.; Haberer,~D.; Fedorov,~A.;
Adamchuk,~V.~K.; Preobrajenski,~A.~B.; Dudin,~P.; Barinov,~A.;
Oehzelt,~M.; {\it et al.}
Nitrogen-Doped Graphene: Efficient Growth, Structure, and
Electronic Properties.
\emph{Nano Lett.} \textbf{2011}, \emph{11}, 5401--5407.

\bibitem{KimN3}
Lin,~T.; Huang,~F.; Liang,~J.; Wang,~Y.
A Facile Preparation Route for Boron-Doped Graphene, and its
CdTe Solar Cell Application.
\emph{Energy Environ. Sci.} \textbf{2011}, \emph{4}, 862--865.

\bibitem{SIESTA} 
Soler,~J.~M.; Artacho,~E.; Gale,~J.~D.; Garc\'ia,~A.; Junquera,~J.;
Ordej\'on,~P.; S\'anchez-Portal,~D. 
The SIESTA Method for {\it Ab Initio} Order-$N$ Materials Simulation.
\emph{J. Phys.: Condens. Matter} \textbf{2002}, \emph{14}, 2745--2779.

\bibitem{Adessi}
Adessi,~Ch.; Roche,~S.; Blase,~X.
Reduced Backscattering in Potassium-Doped Nanotubes: {\it Ab Initio}
and Semiempirical Simulations.
\emph{Phys. Rev. B} \textbf{2006}, \emph{73}, 125414.

\bibitem{Stern}
Stern,~F. Elementary Theory of the Optical Properties of Solids.
In \emph{Solid State Physics. Advances in Research and Applications};
Seitz~F., Turnbull~D., Eds.; Academic Press: New York, 1963;
Vol.~15, pp~299--408.

\bibitem{Son} 
Son,~Y.-W.; Cohen,~M.~L.; Louie,~S.~G.
Energy Gaps in Graphene Nanoribbons.
\emph{Phys. Rev. Lett.} \textbf{2006}, \emph{97}, 216803.

\bibitem{Cohen1} 
Choi,~H.~J.; Ihm,~J.; Louie,~S.~G.; Cohen,~M.~L.
Defects, Quasibound States, and Quantum Conductance in Metallic
Carbon Nanotubes.
\emph{Phys. Rev. Lett.} \textbf{2000}, \emph{84}, 2917--2920.

\bibitem{Cohen2}
Avriller,~R.; Roche,~S.; Triozon,~F.; Blase,~X.; Latil,~S.
Low-Dimensional Quantum Transport Properties of Chemically-Disordered
Carbon Nanotubes: from Weak to Strong Localization Regimes.
\emph{Mod. Phys. Lett. B} \textbf{2007}, \emph{21}, 1955--1982.

\bibitem{hossain_apl2009}
Hossain,~M.~Z.
Chemistry at the Graphene-SiO2 Interface.
\emph{Appl. Phys. Lett.} \textbf{2009}, \emph{95}, 143125.

\bibitem{NanoTCAD} 
Fiori,~G.; Iannaccone,~G. 
Simulation of Graphene Nanoribbon Field-Effect Transistors.
\emph{IEEE Electron Device Lett.} \textbf{2007}, \emph{28}, 760--762.
Code is available at ``NanoTCAD ViDES,''
DOI: 10254/nanohub-r5116.5. http://nanohub.org/resources/vides/
(accessed May 28, 2012).

\bibitem{Datta}
Datta,~S. \emph{Quantum transport: Atom to transistor};
Cambridge University Press: Cambridge, United Kingdom, 2005.

\bibitem{Guo} 
Guo,~J.; Datta,~S.; Lundstrom,~M.; Anantram,~M.~P.
Towards Multiscale Modeling of Carbon Nanotube Transistors.
\emph{Int. J. Multiscale Comput. Eng.} \textbf{2004}, \emph{2}, 257--276.

\bibitem{tang} Tang,~Y.-B.; Yin,~L.-C.; Yang,~Y.; Bo,~X.-H.; Cao,~Y.-L.;
Wang,~H.-E.; Zhang,~W.-J.; Bello,~I.; Lee,~S.-T.; Cheng,~H.-M.; {\it et al.}
Tunable Band Gaps and p-Type Transport Properties of Boron-Doped
Graphenes by Controllable Ion Doping Using Reactive Microwave Plasma.
\emph{ACS Nano} \textbf{2012}, \emph{6}, 1970--1978.

\bibitem{Fujita}
Fujita,~M.; Igami,~M.; Nakada,~K.
Lattice Distortion in Nanographite Ribbons.
\emph{J. Phys. Soc. Jpn.} \textbf{1997}, \emph{66}, 1864--1867.

\bibitem{VMD2}
Humphrey,~W.; Dalke,~A.; Schulten,~K.
VMD: Visual Molecular Dynamics.
\emph{J. Molec. Graphics} \textbf{1996}, \emph{14}, 33--38.
See http://www.ks.uiuc.edu/Research/vmd/
(accessed May 28, 2012).

\end{thebibliography}
\end{document}